\documentstyle[]{article}
\begin{document}
\begin{center}
{ \Huge \bf
On Mechanical Response of \\ Macro- and Microfiller-Reinforced Polymer:\\ Young Modulus and Shear Wave Velocity.}
\end{center}
\vskip1cm
\begin{center}
{\huge   M. Hudak{\footnote{\Large Stierova 23, Kosice} } \\
\vskip0.5cm
O. Hudak{\footnote{\Large Department of Aerodynamics and Simulations, Faculty of Aviation, Technical University Kosice, Kosice} } 
}
\end{center}

\newpage
\section*{Abstract}

The problem of effective shear and bulk moduli, of an effective Poison's ratio and of an effective dielectric response in microcomposites of the ferroelastic-dielectric type was studied
by us recently. Recently one of us, M. Hudak, studied in \cite{MHDP} static mechanical properties of the EVA material theoretically and experimentally.  We extend our mentioned study of mechanical response of microcomposites of the type ferrolastics - dielectrics to study of Young modulus and shear wave velocity.

\newpage
\section{Introduction}

Macro- and microfiller-reinforced polymer composites are used 70 years in various industries \cite{HHIR}. Fiberglass pipe from 1948  can be mentioned. These materials have superior mechanical properties,  they are nonmagnetic and have a good resistance against harsh environment. Thermoplastic and thermosetting composites are used in such sectors as aerospace, automotive sector (molded parts, fuel and gas tanks), rail, construction, sport, medical, electrical, oil/gas, energy and water. Transportation is dominant sector for these composites.  It is possible to obtain high specific stiffness and strength.  Body parts of polymer based composites are in various land transport vehicles. Early use of the composites was also in marine, commercial aircrafts, biomedical applications (hip replacement-low friction coefficient, low density, biocompatibility), boot applications.
  Macro- and microfiller-reinforced polymer composites are dielectric mixtures. Their electrical properties and modeling of dielectric mixtures are discussed in \cite{A01}. They are composites, which are made up of at least two constituents or phases.  The outstanding mechanical properties of them led to extensive research and to a highly developed technologies \cite{A02}, \cite{A03} and \cite{A04}. In \cite{A01} attention was given to their other physical properties, which affected their use in electrical applications \cite{A010}, \cite{A011}, \cite{A012} and \cite{A013}. In \cite{A19} O. Hudak and M. Hudak studied static dielectric response of microcomposites of the type ferrolastics - dielectrics  for application in solid oxide fuel cells (SOFC).

 Recently we have studied dielectric response of ferroelastic-dielectric composites  \cite{HS1} and \cite{HS2}.
When a coupling of the elastic
strain to the electric polarization is present then the dielectric
response of a ferroelastic material may be studied, see in
\cite{jona} and \cite{LG}.

Elastic response function - compliance -
shows in Cole-Cole diagrams circular and non-circular behavior of
these materials in their crystalline and ceramic form \cite{HRS}, thus
multirelaxation phenomena exist in these materials under
some conditions.

Higher mechanical loading leads to 
nonlinear behavior and is exhibited by ferroelectric and ferroelastic
ceramics \cite{E}.

In the paraelastic phase coupling of the elastic strain and electric
polarization does exist at these temperatures. 
Constraint due to neighboring material 
and due to shape of this material inclusions lead, below the critical temperature, for
the transition from paraelastic to ferroelastic phase to several forms of the low-temperature phase
\cite{JCD} in the matrix .  As it is noted in \cite{JCD} only for very
small grains there exists a single variant of this form. We wil further assume
that there are small mechanical fields of the order
$10^{-3}$ and that particles are of the diameter of the order of
1$\mu$m. They may form microcomposites. 
In ferroelastic phase long-range anisotropic forces may appear
\cite{LSRSB}. In our paper we discuss properties of the microcomposite
above the critical temperature, and these long-range forces can be
neglected in the paraelastic phase.

The problem of effective shear and bulk moduli, of an effective Poison's ratio and of an effective dielectric response in microcomposites of the ferroelastic-dielectric type is a very interesting one for the polymers of the EVA type with fillers.  Recently one of us, M. Hudak studied in \cite{MHDP} static mechanical properties of the EVA material theoretically and in experimentally realized form. The Young modulus play also important role. The properties of such materials may be restructured by filers. Then dependence of the Young modulus on the concentration of filler to describe observed dependencies is interesting theoretical problem.
We will use here the model developed in \cite{HS2} to determine the behavior of elastic moduli of the microcomposite. We assume that the grains of both phases are more-less of the spherical shape. The shape of the ferroelastic material particles is assumed to be spherical and characterized by the same diameter, it depends on the method of preparation of the microcomposite. The shape of dielectric particles is assumed to be more-less spherical and we assume further that there exists a distribution of their radii which may be characterized by a homogeneous distribution function as concerning mechanical properties. Analysis of the elastic moduli of heterogeneous materials was done in \cite{B}. We used from \cite{B} method to analyze our system of the ferroelastic-dielectric  microcomposite. We will describe mean-field equations for the effective shear and bulk moduli of the composite, and for the effective Poisson's ratio found in \cite{HS2}.  From these quantities we describe the Young modulus and shear wave velocity, which is the aim of this paper.
Note that from the effective shear modulus $G^{*}$ we can calculate the shear wave velocity in homogeneous and isotropic solids (as elastomers are), where there are two kinds
of waves, pressure waves and shear waves. The effective velocity of a shear wave $v^{*}_{s}$ is given by the shear modulus and the the solid’s density.
A  shear wave $v^{*}_{s}$ (or S-wave) is an elastic body wave - not surface wave. The name S-waves is from seismology where they are secondary waves after compression primary P-waves. Besides geotechnical informations they are used in measuring biological tissues, in architecture , in evaluation of liquifaction of solids.

In this paper we study mechanical properties of microcomposites of the type ferrolastics - dielectrics  for application in macro- and microfiller-reinforced polymer composites. In the first part we describe model of ferroelastic-dielectric microcomposites for polymer and  elastic moduli of the macro- and microfiller-reinforced polymer composite is studied. Then the case of no distribution of the form of particles and effective Young modulus are studied and in the next part low concentration of the dielectric material in the composite and its influence on the Young modulus is described.

\section{Model of ferroelastic-dielectric microcomposites and elastic moduli}

For filler-reinforced polymer composite we describe a model of a ferroelastic-dielectric microcomposite. Primary
order parameter in this material is the corresponding component of the elastic
strain tensor. The polarization is in ferroelastics  a secondary order parameter The ferroelectric
state is present due to a coupling between the elastic strain tensor and the polarization. For mechanical response  and for dielectric response changing
the concentration of two types of particles in the
microcomposite the response to external fields changes. For simplicity we consider all particles in the
microcomposite of the same diameter d for the ferroelastic material and of nearly the same diameter for dielectric material.
We will assume that the composite consists of ferroelastic  particles - spheres. The index N is used for this material constants .  There are also dielectric particles - almost spheres - which have different diameters of the other material. We will assume further that there are (N-1) different particles of this second type. Mechanically we have a composite from N types of particles. The concentration of the ferroelastic particles is assumed to be $(1-x)$, and that of the $i=1, ..., N-1$ types of dielectric particles have the same concentration $x_{i}= \frac{x}{N-1}$ . The total concentration of the second material is x. The size distribution of grains of the second material is assumed to be homogeneous. Then mechanically we have a material of the volume V with the N phases. The volume concentration is $ \frac{V_{i}}{V} $ for the i-th particle (mechanically phase, $i = 1, 2, 3, ..., N.$). We assume that the particles of the second material have the same volume $ \frac{Vx}{N-1} $ . To derive effective moduli of the composite we wil use the method according to \cite{B}. 
We calculated the effective shear modulus $G^{*}$, the effective bulk modulus $K^{*}$ and the effective Poisson´s ratio $\nu^{*}$. In our paper \cite{HS2} we did not calculated the Young modulus and the shear wave velocity. These are other elastic characteristics of the composite. We consider, as in \cite{B},  a large cube of the microcomposite with edges parallel to the coordinate axes (x,y,z). Further we assume that a uniform shear stress may act to the surface of the cube and a uniform hydrostatic pressure may act on the surface of the cube. This lead to a strain and a volume contractions. Note that for large volume of the particles of the second material in the ferroelastic material they will appear as inclusions in a ferroelastic matrix. In \cite{B} the composite is described as a matrix and this includes interactions of the particles of the dielectric material in the ferroelastic material. Thus we obtain effective elastic moduli.

Let us now describe equations for the effective elastic moduli of the microcomposite in the limit of a large number $(N-1)$ of dielectric particles with their concentration $x$.
The effective shear modulus $G^{*}$ is found from the equation $2.7$ in \cite{B}, see also \cite{HS2}:

\begin{equation}
\label{1.}
\frac{1}{G^{*}} = \frac{1}{G_{N}} + x \int^{1}_{0} dr (1 - \frac{G(r)}{G_{N}}) \frac{1}{G^{*} + \beta^{*}(G(r) - G^{*})}
\end{equation}

where:

\begin{equation}
\label{2.}
\beta* = \frac{2(4 - 5 \nu^{*})}{15(1 - \nu^{*})}
\end{equation}

and $ \nu^{*}$ denotes the effective Poisson´s ratio of the composite material.

From the effective shear modulus $G^{*}$ we can calculate the shear wave velocity in homogeneous and isotropic solids (as elastomers are), where there are two kinds
of waves, pressure waves and shear waves. The effective velocity of a shear wave $v^{*}_{s}$ is given by the shear modulus and the the solid’s density:

\begin{equation}
\label{2..}
v^{*}_{s}= \sqrt{\frac{G^{*}}{\rho}}
\end{equation}

where $G^{*}$ is the shear modulus and $\rho$ is the solid’s density.

The effective bulk modulus $K^{*}$ is found from the equation:

\begin{equation}
\label{3.}
\frac{1}{K^{*}} = \frac{1}{K_{N}} + x \int^{1}_{0} dr (1 - \frac{K(r)}{K_{N}}) \frac{1}{K^{*} + \alpha^{*}(K(r) - K^{*})}
\end{equation}

where:

\begin{equation}
\label{4.}
\alpha^{*} = \frac{1 +  \nu^{*}}{3(1 - \nu^{*})}.
\end{equation}

Here $ \nu^{*}$ ,  the effective Poisson´s ratio of the composite material:

\begin{equation}
\label{5.}
\nu^{*} = \frac{3 K^{*} -  2 G^{*}}{6K^{*} + 2 G^{*}}
\end{equation}

Thus equations (\ref{1.}) - (\ref{5.}) give the dependence of moduli $ G^{*} $  and $ K^{*} $ on the elastic constants $G_{N}$ and $K_{N}$, and on $G(r)$ and $K(r)$  distribution functions of shear and bulk moduli for the second material. Here r denotes different types of particles of the second material.

Let us assume that the distribution $G(r)$ of the elastic moduli for particles of the second material is linear:

\begin{equation}
\label{6.}
G(r) = r G_{1} + (1 - r) G_{0}
\end{equation}

and:

\begin{equation}
\label{6..}
K(r) = r K_{1} + (1 - r) K_{0}
\end{equation}

where $G_{0}$ and $G_{1}$ are two values of the shear modulus G for the second material which correspond to the index 0 and 1 of the particles diameter, and where $K_{0}$ and $K_{1}$ are two values of the bulk modulus K for the second material which correspond to the index 0 and 1 of the particles diameter. Let us note that the linear approximation is expected to be good approximation for the case in which values for the moduli for $r =  0$ and for $r = 1$ are not different too much. The larger difference $ G_{1} - G_{0}$ for the second material is obtained  due to presence of texture, cracks or other types of defects.

From (\ref{6.}) - (\ref{6..}) we obtain equations (\ref{1.}) - (\ref{5.}) in the following form.
The effective shear modulus $G^{*}$ is found from the equation (\ref{1.}) where $G(r)=G$:

\begin{equation}
\label{7.}
\frac{1}{G^{*}} = \frac{1}{G_{N}} + \frac{x}{G_{N} \beta^{*}} [\frac{\beta^{*}(G^{*} - G_{N}) - G^{*}}{\beta^{*} (G_{0} - G_{1})} \ln (\frac{G^{*} + \beta^{*}(G_{1} - G^{*})}{G^{*} + \beta^{*}(G_{0} - G^{*})}) - 1]
\end{equation}

where $ \beta^{*}$ is given by (\ref{2.})
and $ \nu^{*}$ is the Poisson´s ratio of the composite material.

From the effective shear modulus we can calculate the shear wave velocity in homogeneous and isotropic solids. The effective velocity of a shear wave $v^{*}_{s}$ is given by the shear modulus and the solid’s density in this case as:

\begin{equation}
\label{2...}
\frac{1}{v^{*}_{s}}= \sqrt{\frac{\rho}{G^{*}}}= \sqrt{\rho .(\frac{1}{G_{N}} + \frac{x}{G_{N} \beta^{*}} [\frac{\beta^{*}(G^{*} - G_{N}) - G^{*}}{\beta^{*} (G_{0} - G_{1})} \ln (\frac{G^{*} + \beta^{*}(G_{1} - G^{*})}{G^{*} + \beta^{*}(G_{0} - G^{*})}) - 1]) }.
\end{equation}

The effective bulk modulus $K^{*}$ is found from the equation:

\begin{equation}
\label{9.}
\frac{1}{K^{*}} = \frac{1}{K_{N}} + \frac{x}{K_{N}\alpha^{*}} [\frac{\alpha^{*} (K^{*} - K_{N}) - K^{*}}{\alpha^{*}(K_{0} - K_{1})} \ln (\frac{K^{*} + \alpha^{*}(K_{1} - K^{*})}{K^{*} + \alpha^{*}(K_{0} - K^{*})}) - 1]
\end{equation}

where $ \alpha^{*}$ is given by (\ref{4.}).

From the equations (\ref{7.}) - (\ref{9.}) it follows that such textures, cracks and defects which induce larger difference of the elastic moduli G and K lead in the first approximation of the inverse of the absolute value of the difference $ G_{1} - G_{0}$ and of the difference $ K_{1} - K_{0}$ to the microcomposite in which:

\begin{equation}
\label{9... }
\frac{1}{K^{*}} \approx \frac{1}{K_{N}} (1 - \frac{x}{\alpha^{*}})
\end{equation}

and:

\begin{equation}
\label{9.... }
\frac{1}{G^{*}} \approx \frac{1}{G_{N}} (1 - \frac{x}{\beta^{*}})
\end{equation}

and where for $\alpha^{*} $ and $ \beta^{*} $ positive hardening of the microcomposite is present in this order. This is observed in numerical simulation,  \cite{GCL}, where crystallographic microstructure is taken into account in the numerical modeling.

The equations (\ref{7.}) - (\ref{9.}) have different form for the case in which there is no distribution of the elastic moduli of particles of the second material, e.i. when $ G_{0} = G_{1} = G $ and when $ K_{0} = K_{1} = K $. Then we obtain that the equations (\ref{7.}) - (\ref{9.}) take the following form.

The effective shear modulus $G^{*}$ is found from the equation:

\begin{equation}
\label{12.}
\frac{1}{G^{*}} = \frac{1}{G_{N}} - \frac{x}{G_{N} } [\frac{G - G_{N}}{G^{*} + \beta^{*}(G - G^{*})} ]
\end{equation}

From the effective shear modulus (\ref{12.}) we can calculate the shear wave velocity . The effective velocity of a shear wave $v^{*}_{s}$ is given now by:

\begin{equation}
\label{2....}
\frac{1}{v^{*}_{s}}= \sqrt{\rho. ( \frac{1}{G_{N}} - \frac{x}{G_{N} } [\frac{G - G_{N}}{G^{*} + \beta^{*}(G - G^{*})} ])}.
\end{equation}

and the effective bulk modulus $K^{*}$ is found from the equation (\ref{3.}) where $K(r)=K$:

\begin{equation}
\label{14.}
\frac{1}{K^{*}} = \frac{1}{K_{N}} - \frac{x}{K_{N}} [\frac{K - K_{N}}{K^{*} + \alpha^{*}(K - K^{*})} ].
\end{equation}

Young modulus $E^{*}$ for isotropic composite may be calculated  \cite{LL}:

\begin{equation}
\label{15.}
E^{*}= 3.K^{*}(1-2.\nu^{*}) = 3.\frac{1}{\frac{1}{K_{N}} - \frac{x}{K_{N}} [\frac{K - K_{N}}{K^{*} + \alpha^{*}(K - K^{*})} ]}(1-2.\nu^{*})] = 
\end{equation}
\[= 2.G^{*}(1+\nu^{*}) = 2.\frac{1}{ \frac{1}{G_{N}} - \frac{x}{G_{N} } [\frac{G - G_{N}}{G^{*} + \beta^{*}(G - G^{*})} ]}(1+\nu^{*})] \]

Here the Poison´s ratio $\nu^{*}$ we obtain from above. To obtain the Young modulus $E^{*}$ in the first order of x we may omit stars in $K^{*}$, $\alpha^{*}$, $\nu^{*}$, $\beta^{*}$ and $ G^{*}$ in the equation (\ref{15.}). In this chapter we have formulated a model theory for the elastic moduli G, K and Y of a microcomposite which has two constituents and for the effective shear wave velocity.

\section{The case of no distribution of the form of particles and effective elastic moduli}

For the filler-reinforced polymer composite with no distribution of the form of particles in (\ref{12.}) - (\ref{14.}) we have described in \cite{HS2} the equations for the effective shear and bulk moduli. To study the solution of these equations it is convenient to introduce the following ansatz in the equations (\ref{12.}) and (\ref{14.}):

\begin{equation}
\label{17..}
G^{*} = (G - G_{N}).g \\
\end{equation}
\[ K^{*} = (K - K_{N}).k \]

The functions g and k are given by:

\begin{equation}
\label{18..}
g_{\pm} = \frac{1}{2[1 - \beta^{*}]} [- \beta^{*}(\gamma + \gamma_{N})  +  (x + \gamma_{N}) \pm \sqrt{D_{g}}]
\end{equation}

where:

\begin{equation}
\label{19..}
D_{g} = [\beta^{*}(\gamma  + \gamma_{N})  - (x + \gamma_{N})]^{2} + 4 (1 - \beta^{*})\beta^{*} \gamma \gamma_{N}
\end{equation}

and where:

\begin{equation}
\label{20.}
\gamma_{N} = \frac{G_{N}}{G - G_{N}}
\end{equation}

and:

\begin{equation}
\label{21.}
\gamma = \frac{G}{G - G_{N}},
\end{equation}
 
and:

\begin{equation}
\label{22.}
k_{\pm} = \frac{1}{2[1 - \alpha^{*}]} [- \alpha^{*}(\delta + \delta_{N})  +  (x + \delta_{N}) \pm \sqrt{D_{k}}]
\end{equation}

where:

\begin{equation}
\label{23.}
D_{k} = [\alpha^{*}(\delta  + \delta_{N})  - (x + \delta_{N})]^{2} + 4 (1 - \alpha^{*})\alpha^{*} \delta \delta_{N}
\end{equation}

and where:

\begin{equation}
\label{24.}
\delta_{N} = \frac{K_{N}}{K - K_{N}}
\end{equation}

and:

\begin{equation}
\label{25.}
\delta = \frac{K}{K - K_{N}}.
\end{equation}

The effective Poison's ratio $\nu^{*}$ has the form:

\begin{equation}
\label{26.}
\nu^{*} = \frac{3k - 2g \epsilon}{6k + 2g \epsilon}
\end{equation}

where $\epsilon $ is defined as:

\begin{equation}
\label{27.}
\epsilon = \frac{G - G_{N}}{K - K_{N}}.
\end{equation}

Note that $D_{g}$ and $D_{k}$ are discriminants which may be positive or negative. In the first case $g_{\pm}$ is real and $k_{\pm}$ is real too.
In the second case $g_{\pm}$ is a complex number  and $k_{\pm}$ is a complex number too. The signs of both discriminants in equations (\ref{18.}) and (\ref{20.}) are independent.
We will consider the static case only in the following. According to the fluctuation-dissipation theorem the imaginary part of the moduli will be zero. 
For isotropic elastic solids \cite{TG} the range of stability for $\nu^{*} $ is between -1 and 0.5 values. We will assume that the microcomposite is an isotropic elastic solid. The microcomposite is assumed to be unconstrained. The range of the Poison's ration $ \nu^{*} $ above corresponds to the stability criteria, e.i. to positive $G^{*}$ and $K^{*}$ mechanical  moduli. Ferroelastic materials may have a negative stiffness, see in \cite{S}. The problem with negative stiffness inclusions /clusters/ in a matrix may be considered by our model, this problem is not discussed here.  Note that in \cite{LET} ferroelastic martensites parent to product phase transition was studied. Order parameter which drives this transition is shear deformation. Strain energy dominates morphology, and the transition results in a characteristic lamellar or twinned structure. This then may lead to a nonisotropic  microcomposite, and calculations of the mechanical moduli should take this fact into account. This is not the case in this paper for simplicity. Martensites are proper ferroelastics, and improper ferroelastics are f.e. ferroelectrics, the martensites show weakly first order phase transition near the second order, or second order phase transition and thus Landau theory applies. This fact is used in our paper. Examples of proper ferroelastics are InTl, FePd, NiTi, AuCd. In proper ferroelastics textures form occurs and it does not correspond with the assumption of the isotropic elastic microcomposite. Improper ferroelastics are materials like high-temperature superconductors, and GMR manganites. In \cite{WL} it was studied how composites with inclusions of negative bulk modulus in a matrix behave.  Let us note that complex values of mechanical moduli correspond to viscoelastic materials, in which however $G^{''} $, the imaginary part of the shear modulus, must be positive \cite{CH} if the material should be stable.
The quantity $\tan(\delta)$ is given by the ratio $ \frac{G^{''}}{G^{'}}$, where $ G^{'}$ is the real part of the shear modulus. Experimental results for this ratio are from 0.01 at room temperature  to 0.015 at higher temperatures. In \cite{JL} it was found that ripples are present in the response for temperature dependence of the 5-percent  $VO_{2} $ in Sn matrix.

We can calculate the shear wave velocity in homogeneous and isotropic solids  where there are two kinds
of waves, pressure waves and shear waves. The effective velocity of a shear wave $v^{*}_{s}$ is given by the shear modulus from (\ref{17.}) :

\begin{equation}
\label{2......}
v^{*}_{s}= \sqrt{\frac{(G - G_{N}).g}{\rho}}
\end{equation}

where $(G - G_{N}).g$ are given above.

The Young modulus $E^{*}$ for isotropic composite may in this case be calculated  from:

\begin{equation}
\label{15,.}
E^{*}= 2.G^{*}(1+\nu^{*}) = 2. (G - G_{N}).g .(1+\nu^{*})=3.G^{*}(1-2\nu^{*}) =3.(K-K_{N}).k.(1-2\nu^{*}) 
\end{equation}

where $\nu^{*}$ is given from (\ref{17.}).

\section{Low concentration of the dielectric material: elastic moduli}

In Macro- and Microfiller-Reinforced Polymer Composite with low concentration of the dielectric material we obtain from the equations (\ref{12.}) - (\ref{14.}) in the limit of small concentrations x of the dielectric particles the effective Poison's ratio:

\begin{equation}
\label{17.}
\nu^{*} = \frac{3 K_{N} -  2 G_{N}}{6K_{N} + 2 G_{N}}(1 - x \theta)
\end{equation}

where the constant $ \theta $ is given by:

\begin{equation}
\label{17....}
\theta \equiv (\frac{\frac{3K_{N}}{\delta_{N} + \alpha^{*} \delta} - \frac{2G_{N}}{\gamma_{N} + \beta^{*} \gamma}}{3 K_{N} -  2 G_{N}} - \frac{\frac{6K_{N}}{\delta_{N} + \alpha^{*} \delta} + \frac{2G_{N}}{\gamma_{N} + \beta^{*} \gamma}}{6K_{N} + 2 G_{N}})
\end{equation}

and the effective bulk modulus:

\begin{equation}
\label{18.}
K^{*} = K_{N}(1  - \frac{x}{\delta_{N} + \alpha^{*} \delta})
\end{equation}

and the effective shear modulus:

\begin{equation}
\label{19.}
G^{*} = G_{N}(1 - \frac{x}{\gamma_{N} + \beta^{*} \gamma})
\end{equation}

where $\alpha$ and $\beta$ constants are calculated as above.
We have found that:

\begin{equation}
\label{19...}
\beta^{*} = \beta_{N}(1 + x ( \frac{5 \nu_{N}\theta}{2 (4 - 5\nu_{N})} -  \frac{ \nu_{N}\theta}{15 (1 - \nu_{N})}))
\end{equation}

and for $ \alpha^{*}$ we have found:

\begin{equation}
\label{19....}
\alpha^{*} = \alpha_{N}(1 - x ( \frac{ \nu_{N}\theta}{3 (1 + \nu_{N})} +  \frac{ \nu_{N}\theta}{3 (1 - \nu_{N})}))
\end{equation}

Note however that in the equations (\ref{18.}) - (\ref{19.}) we can omit the star at the quantities $\alpha$ and $\beta$ due to the fact that the calculations are done up to the first order in the concentration x.
From the equations (\ref{18.}) we can find that the material with inclusions is less harder if the microcomposites is such that the quantity $ \delta_{N} + \alpha_{N}\delta $ is positive. And vice versa. From the equation  (\ref{19.}) we find that the material with inclusions stiffens if the microcomposite is such that the inequality $ \gamma_{N} + \alpha_{N}\gamma $ is positive. And vice versa.

The shear wave velocity $v^{*}_{s}$ is given by:

\begin{equation}
\label{2.......}
v^{*}_{s}= \sqrt{\frac{G_{N}(1 - \frac{x}{\gamma_{N} + \beta^{*} \gamma})}{\rho}}.
\end{equation}

With increasing concentration x of inclusion stiffening the composite the shear wave velocity decreases.

The Young modulus $E^{*}$ in this case is calculated  to be:

\begin{equation}
\label{15..}
E^{*}= 2.G^{*}(1+\nu^{*}) = 2. G_{N}(1 - \frac{x}{\gamma_{N} + \beta^{*} \gamma}) .(1+\frac{3 K_{N} -  2 G_{N}}{6K_{N} + 2 G_{N}}(1 - x \theta))
\end{equation}

This calculation is to the $O(x)$ order so we neglect in (\ref{15.}) the term of the order $O(x^{2})$:

\begin{equation}
\label{15...}
E^{*}= 2.G^{*}(1+\nu^{*}) = 2. G_{N}[(1+\frac{3 K_{N} -  2 G_{N}}{6K_{N} + 2 G_{N}}) -
x. (\frac{1}{\gamma_{N}+ \beta^{*} \gamma} - \theta \frac{3 K_{N} -  2 G_{N}}{6K_{N} + 2 G_{N}})]
\end{equation}

We ommited a star in $\beta^{*}$ in the equation (\ref{15..}) in the $O(x)$ order.

\section{Summary}
According to the NASA Survey of Breakthrough Materials \cite{CC} applications of new materials must be evaluated in a systems context.  Advanced structural design methods and highly efficient structural
concepts are  required to exploit the potential benefits of biomimetic, nanostructured, multifunctional materials in revolutionary aerospace vehicles.
Note that for structural materials for vehicles and habitats a factor of 2 gain in weight savings can be achieved by carbon fiber reinforced polymers, metal matrix
composites, and intermetallics;  carbon nanotube reinforced polymers (and metals) may gain a factor of 10 gain in weight savings.
For structural materials for propulsion components ceramics may gain a factor of 2 gain in use temperature but may never achieve attractive structural design
allowables; advanced metallic alloys and intermetallics may offer a factor of 2 gain in weight savings but only modest temperature improvements; polymer matrix
composites, including carbon nanotubes, may offer significant weight savings but at a reduction in the use temperature.
Materials for radiation shielding may gain by selecting structural materials only modest improvement in shielding potential $(10-20\%)$; dramatic
improvements in radiation protection may be achieved by nonconventional vehicle and habitat configurations.
In thermal protection systems breakthroughs will not come from improved material properties but from revolutionary concepts and capabilities such as sharp
leading edges, rapid heat transfer, all-weather durability, self-diagnostics and self-repair.   In electronic and photonic  materials dramatic breakthroughs will occur from
functionalized nanostructured materials enabling the fabrication of nano-electro-mechanical systems (NEMS).
As we can see there are many possibilities to gain in weight mainly for structural materials for vehicles - reinforced polymers, structural materials for propulsion components -   polymer matrix component, materials for radiationshielding, thermal protection and electronic and photonic materials.

As we mentioned above macro- and microfiller-reinforced polymer composites are used 70 years in various industries,  \cite{HHIR}, these materials have superior mechanical properties,  they are nonmagnetic and have a good resistance against harsh environment. Vehicles are lighter, costs reduced, they are corrosion resistant, body parts of polymer based composites are in various land transport vehicles. Use of the composites was also in marine, commercial aircrafts, biomedical applications (hip replacemment-low friction coefficient, low density, biocompatibility), boot applications.

It is possible to obtain high specific stiffness and strength and gains for these materials using our calculations of the bulk and shear moduli in \cite{HS2} and using our calculations in this paper for the Young modulus and for the shear wave velocity. Our calculations for material design may be used not only in the first approximation in the concentration $x$  of the dielectric material, but also higher mechanical loading which leads to nonlinear behavior and is exhibited by ferroelectric and ferroelastic ceramics \cite{E} may be studied.

The problem of effective shear and bulk moduli, of an effective Poison's ratio and of an effective dielectric response in microcomposites of the ferroelastic-dielectric type was studied recently, \cite{HS1} and \cite{HS2}, as well as one of us, M. Hudak studied in \cite{MHDP} static mechanical properties of the EVA material theoretically and in experimentally.  Thus it was useful to extend our study of mechanical response of microcomposites of the type ferrolastics - dielectrics to study the shear wave velocity and to  study the Young modulus. We used in this paper our \cite{HS2} model of ferroelastic-dielectric microcomposites and elastic moduli. Then we studied the case of no distribution of the form of particles and effective Young moduli and effective shear wave velocity. Low concentration of the dielectric material was then studied for the Young modulus and the shear wave velocities.


\begin{thebibliography}{999999}
\bibitem{MHDP} 
M. Hudak, Diploma Work,  University of Zilina, Zilina
\bibitem{HHIR}
Sabu Thomas (Editor), Kuruvilla Joseph (Editor), S. K. Malhotra (Editor), Koichi Goda (Editor), M. S. Sreekala (Editor), Polymer Composites, Volume 1, Macro- and Microcomposites, ISBN: 978-3-527-32624-2, 814 pages, March 2012, John Wiley and Sons Ltd, London, chap. 23. Applications of Macro- and Microfiller-Reinforced Polymer Composites.
\bibitem{A01}
Enis Tuncer, Yuriy V. Serdyuk and Stanislaw M. Gubanski http://arxiv.org/abs/cond-mat/0111254v2
\bibitem{A02}
D. Hull and T. W. Clyne. An introduction to composite materials. Cambridge Solid Sate
Science Series. Cambridge University Press, Cambridge, second edition, 1996
\bibitem{A03}
E. W. MacFarland and W. H. Weinberg. Combinatorial approaches to materials discovery.
TIBTECH, 17:107–115, March 1999
\bibitem{A04}
P. Gilormini and Y. , Materials Science and Engineering, 7:805–816, 1999

\bibitem{A010}
D. K. Hale.\,   Journal of Materials Science, 11:
2105–2141, 1976.

\bibitem{A011}
R. Landauer. Electrical conductivity in inhomogeneous media. In J. C. Garland and D. B.
Tanner, editors, Electrical Transport and Optical properties of Inhomogeneous Media,
volume 40 of AIP Conference Proceedings, pages 2–43, New York, 1978. American Institute
of Physics.
\bibitem{A012}
F. Lux. ,  Journal of Materials Science, 28:285–301, 1993
\bibitem{A013}
Ari Sihvola. Electromagnetic mixing formulas and applications, volume 47 of IEE
Electromagnetic Waves Series. The Institute of Electrical Engineers, London, 1999


\bibitem{A19}
O. Hudak and M. Hudak, arXiv:1610.09617 2016

\bibitem{HS1}O. Hudak and W. Schranz, http://arxiv.org/cond-mat/0511704 , 2005 

\bibitem{HS2}    O. Hudak, W. Schranz,  arXiv:cond-mat/0604590 2006


\bibitem{jona}
F. Jona and G. Shirane: {\it Ferroelectric Crystals}, Dover, New
York, 1962

\bibitem{LG}M.E. Lines and A.M. Glass: {\it Principles and Application
of Ferroelectrics and Related Materials}, Clarendon Press, Oxford,
1977


\bibitem{HRS}R.J. Harrison, S.A.T. Redfern and E.K.H. Salje,
  Phys.Rev. {\bf B 69} (2004) 144101

\bibitem{E}M. Elhadrouz, J. Intell. Mat.Sys.Struct., {\bf 16} (2005)
  221-236

\bibitem{JCD}A.E. Jacobs, S.H. Curnoe and R.C. Desai, Phys.Rev. {\bf
  B68} (2003) 224104


\bibitem{LSRSB}T. Lookman, S.R. Shenoy, K.O. Rasmussen, A. Saxena and
  A.R. Bishop, http://arxiv.org/cond-matt/0211425 , 2002



\bibitem{B}B. Budianksy, J.Mech.Phys. Solids, {\bf 13   } (1965) 223 -
  227

\bibitem{GCL} R.E. Garcia, W.C. Carter and S.A. Langer, J.Am.Soc. {\bf 88} (2005) 750 - 757

\bibitem{LL} L.D. Landau and E.M. Lifshitz, Theory of Elasticity, Vol. 7, p. 13, Pergamon Press, Oxford, 1970

\bibitem{TG} S.P. Timoshenko and J.N. Goodier, {\it Theory of Elasticity}, 3rd ed., McGraw-Hill, New York, 1970

\bibitem{S} E.K.H. Salje, {\it Phase Transitions in Ferroelastic and Co-elastic Crystals}, Cambridge University Press, Cambridge, 1990

\bibitem{LET}T. Lookman et al., http://arxiv.org/cond-mat/02114525 , 2002

\bibitem{WL}Y.C. Wang and R.S. Lakes, J. Composite Materials {\bf 39} (2005) 1645 - 1657

\bibitem{CH}R.M. Christensen, Trans.Soc. Rheology {\bf 16} (1972) 603


\bibitem{JL} T. Jaglinski and R.S. Lakes, Philosophical Magazine Letters {\bf 84} (2004) 803 - 810

\bibitem{CC} Charles E. Harris The Center of Excellence for Structures and Materials NASA Langley Research Center
July 2000

\end{thebibliography}
\end{document}